\documentclass[letter]{aa}  

\usepackage{graphicx}
\usepackage{txfonts}
\usepackage[version=4]{mhchem}
\newcommand{\edt}[1]{{#1}}

\newcommand{\ignore}[1]{}

\begin{document}

    \title{Jupiter formed as a pebble pile around the \ce{N2} ice line}

   \author{A. D. Bosman
          \inst{1}\thanks{\emph{Present address:} Department of Astronomy, University of Michigan, 311 West Hall, 1085 S. University Avenue, Ann Arbor, MI 48109, USA}
          \and
          A. J. Cridland \inst{1}
          \and
          Y. Miguel \inst{1}
          }

   \institute{
   Leiden Observatory, Leiden University, PO Box 9513, 2300 RA Leiden, The Netherlands \\ \email{arbos@umich.edu}}
   
   \date{\today}

 
  \abstract
 {The region around the \ce{H2O} ice line, due to its higher surface density, seems to be the ideal location to form planets. The core of Jupiter, as well as the cores of close in gas giants are thus thought to form in this region of the disk. Actually constraining the formation location of individual planets has proven to be difficult, however.   } 
   {We aim to use the Nitrogen abundance in Jupiter, which is around 4 times solar, in combination with \textit{Juno} constraints on the total mass of heavy elements in Jupiter, to narrow down its formation scenario.}
   {Different pathways of enrichment of Jupiter's atmosphere, \edt{such as the accretion of enriched gas, pebbles or planetesimals} are considered and their implications for the oxygen abundance of Jupiter is discussed.} 
   {The super solar Nitrogen abundance in Jupiter necessitates the accretion of extra \ce{N2} from the proto-solar nebula. The only location of the disk that this can happen is outside, or just inside the \ce{N2} ice line. These constraints favor a pebble accretion origin of Jupiter, both from the composition as well as from a planet formation perspective. We predict that Jupiter's oxygen abundance is between 3.6 and 4.5 times solar. }
   {}


   \maketitle
%

\section{Introduction}
There are currently three theories dealing with the formation of gas giants in the solar system. The classical picture is the core accretion scenario, in which km-size planetesimals grow through mutual collisions. Eventually leading to a core of a few Earth masses which can start to efficiently capture a gaseous atmosphere. \citep{Pollack1996,KokuboIda2002,IdaLin2004}.  For planetesimal accretion to work on a reasonable timescale, within the few Myr lifetime of the gas disk, high surface densities of planetesimals are needed. As such planetesimal accretion is most efficient in forming giant planets at small radii. An increase in the surface density of planetesimals at the \ce{H2O} ice line at a few AU makes this the preferred location for the formation of Jupiter in this scenario \citep[][]{Stevenson1988,Ciesla2006, Schoonenberg2017}. Further migration, due to interactions with the gas disk and resonances with Saturn would then put Jupiter at its current location \citep{Walsh2011}. 

An alternative to the model of core accretion is the paradigm of pebble accretion, in which planetesimals grow by accreting millimeter and centimeter sized pebbles that flow radially through the disk. As the planet migrates as it is accreting its gas, the cores of these planets need to form at larger radii, outside of 15 AU, to end up at a few AU when the gas disk has dissipated \citep[][ Cridland et al. subm.]{Bitsch2015, Bitsch2019}. Finally there is the possibility of forming giant planets through gravitational instabilities in the outer disk. In this scenario, Jupiter would form by direct gravitational collapse of a clump in the cold outer regions of the solar nebula, before migrating in \citep[][]{Boss1997, Boss2002, Boley2010}. 

This sets up a dichotomy of the origin of Jupiter, either formation around the water ice line, or formation at large radii. These different formation histories should leave an imprint on the composition of the planet, especially on the C/O ratio \citep{Oberg2011}. While a lot of effort has been put into constraining composition of Jupiter's atmosphere \citep{Atreya2003, Atreya2016, Bolton2017}, there is a consensus that the oxygen abundance measurement by the \textit{Galileo} mission is not representative for the bulk atmospheric abundance of oxygen in Jupiter, and thus the C/O ratio cannot be used \citep{Niemann1998, Atreya2003}. These studies have found however that both carbon and nitrogen are enhanced above solar levels \citep[see Table~\ref{tab:abundances},][]{Asplund2009, Atreya2016}. The enhancement of nitrogen is interesting, as nitrogen in the interstellar medium (ISM) is extremely volatile, since the main carrier, \ce{N2}, does not freeze-out until temperatures below $\sim$ 20 K are reached \citep{Bisschop2006}. Furthermore, the next most abundant nitrogen carrier, \ce{NH3}, generally does not contain more than 10\% of the total nitrogen budget \citep{Lodders2009, Boogert2015, Cleeves2018, Pontoppidan2019, Altwegg2019}. This potentially makes the nitrogen content of a planet a powerful probe of its formation location. In this letter we use recent insights in planet formation theory and new constraints from the \textit{Juno} mission to put Jupiter in an astrochemical context. This approach brings forward a couple of formation scenarios for Jupiter that can explain the abundance of elemental \ce{N} in its atmosphere. 
\begin{table}
    \centering
    \caption{\label{tab:abundances}Elemental abundances relative to H}
    \begin{tabular}{c c c c}
    \hline 
    \hline
    Element & Protosolar & Jupiter & Enhancement\\ 
    \hline
    C & $2.95 \times 10^{-4}$ & $1.19 \pm 0.29\times 10^{-3}$ & $4.02 \pm 0.98$ \\ 
    N & $7.41 \times 10^{-5}$ & $3.32 \pm 1.27\times 10^{-4}$ & $4.48 \pm 1.71$ \\ 
    O & $5.37 \times 10^{-4}$ & $2.45 \pm 0.80\times 10^{-4}{^*}$ & $0.46 \pm 0.15^*$  \\ 
    \hline
    \end{tabular}
    \tablefoot{Proto-solar abundances from \citet{Asplund2009}, Jupiter abundances from \citet{Atreya2016}\\
    $^*$Oxygen measurement in a hot spot and probably does not represent bulk oxygen abundance of the atmosphere. }
    
\end{table}


During the development of this work, \citet{Oberg2019} published a similar line of argument as discussed here. We share their conclusion, that Jupiter must have formed outside of the \ce{N2} iceline. Enrichment of Jupiter's atmosphere by \ce{N2} is critical in both works, but whereas \citet{Oberg2019} use the relative enrichment pattern of volatile species to constrain the composition and formation temperature of the enriching bodies, we use the total heavy element mass to constrain Jupiter's most likely formation location.


\section{Enriching Jupiter with Nitrogen}
\begin{figure}
    \centering
    \includegraphics[width = \hsize]{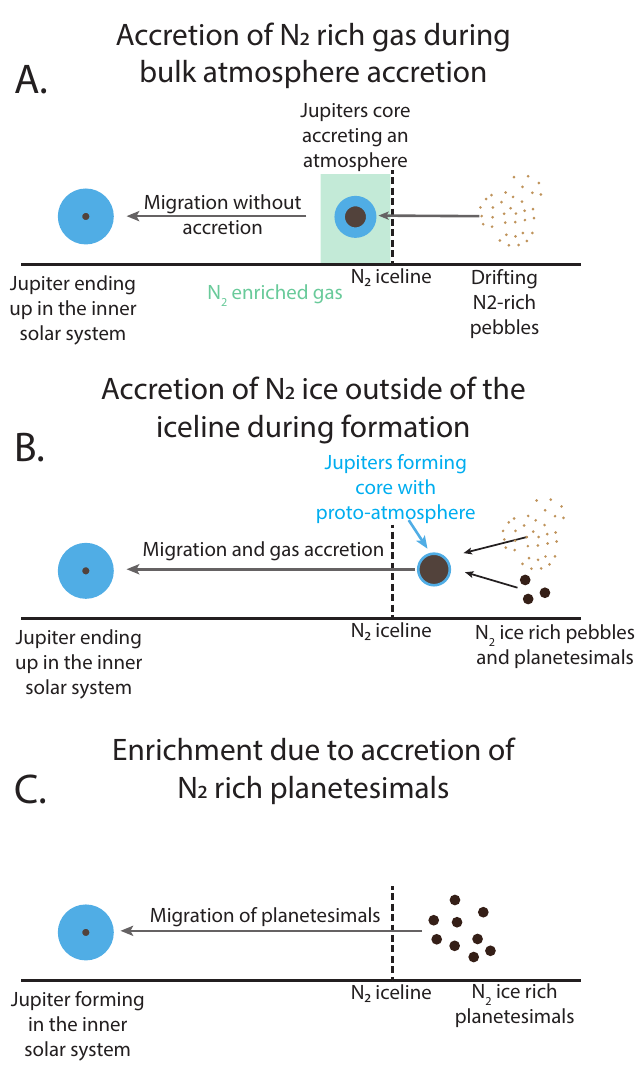}
    \caption{\label{fig:form_scenario} Different scenarios for the origin of nitrogen in Jupiter's atmosphere. 
    In scenario A (\textit{top}), the nitrogen in accreted during the bulk of the atmosphere accretion from the part of the disk that is rich in \ce{N2} gas close to the \ce{N2} ice line. The gas is enriched by rapidly drifting pebbles from outside the \ce{N2} ice line. 
    In scenario B (\textit{middle}), nitrogen is brought in with the solid material that accretes onto Jupiter while it is in the cold outer disk. This limits core formation to outside the \ce{N2} ice line, leaving the location of gas accretion unconstrained. 
    In scenario C (\textit{bottom}), Jupiter forms somewhere inside the \ce{N2} ice line, as far in as the \ce{H2O} ice line, the classical location of Jupiter formation. \ce{N2} then has to be brought in on planetesimals that originate outside of the \ce{N2} ice line and migrate towards the location of the forming Jupiter. This scenario leaves almost no room for other solids than the \ce{N2} rich solids to be accreted by Jupiter after the initial core has formed. \edt{Scenario B seems to be the most reasonable scenario.} }
\end{figure}

Generally speaking there are three methods to enrich the atmosphere of a gas giant planet in a specific element: through the accretion of enriched gas, through the accretion of solids during core or atmosphere formation, or through the late accretion of solids after the planet has accreted its atmosphere (see Fig.~\ref{fig:form_scenario}). Recently there have been multiple studies that have looked at the effect of disk evolution, especially the growth and drift of icy grains, at the effect this has on the gas-phase elemental abundances \citep{Ciesla2006, Booth2017, Stammler2017, Bosman2018, Krijt2018, Booth2018}. In general, it is found that enrichments above solar abundances in a certain element can happen just inside an ice line if radial drift is efficient and the ice line corresponds to a species that is an abundant ($\gtrsim 10\%$) carrier of that element.


In the case of nitrogen, \citet{Booth2018} find an enrichment of elemental nitrogen up to a factor of 2 above solar both within the \ce{NH3} ice line and in an annulus just within the \ce{N2} ice line. The high elemental nitrogen abundances within the \ce{NH3} ice line are strongly dependent on the initial \ce{NH3} abundance. \citet{Booth2018} put 50\% of elemental nitrogen in \ce{NH3}. Such high \ce{NH3} abundances have not been seen in observations of the cold ISM \citep{Boogert2015}, proto-planetary disks \citep{Cleeves2018,Pontoppidan2019} or solar system objects \citep{Lodders2009,Altwegg2019}, indicating that in all these environments \ce{N2} is the dominant carrier, containing $\sim 90\%$ of the elemental nitrogen. As such a super-solar nitrogen abundance due to the sublimation of \ce{NH3} is unlikely. This means that if Jupiter's atmospheric enhancements are due to the accretion of enriched gas (Fig. 1, scenario A), it must have accreted most of its mass just within the \ce{N2} ice line. 

The second and third scenario depend on the enrichment by solids that deposit nitrogen in the atmosphere. For simplicity, we assume that Jupiter accreted a solar N/H ratio from the gas, and that all of the extra nitrogen is brought in by solids. We furthermore assume that the solids that are accreted onto Jupiter deposited their full nitrogen reservoir into the atmosphere. With these assumption it is possible to calculate the mass of refractories (silicates and metals) that Jupiter needs to accrete as a function of the nitrogen to refractory mass ratio for the accreting solid bodies. 


\begin{figure*}
\sidecaption
\centering
\includegraphics[width = 12cm]{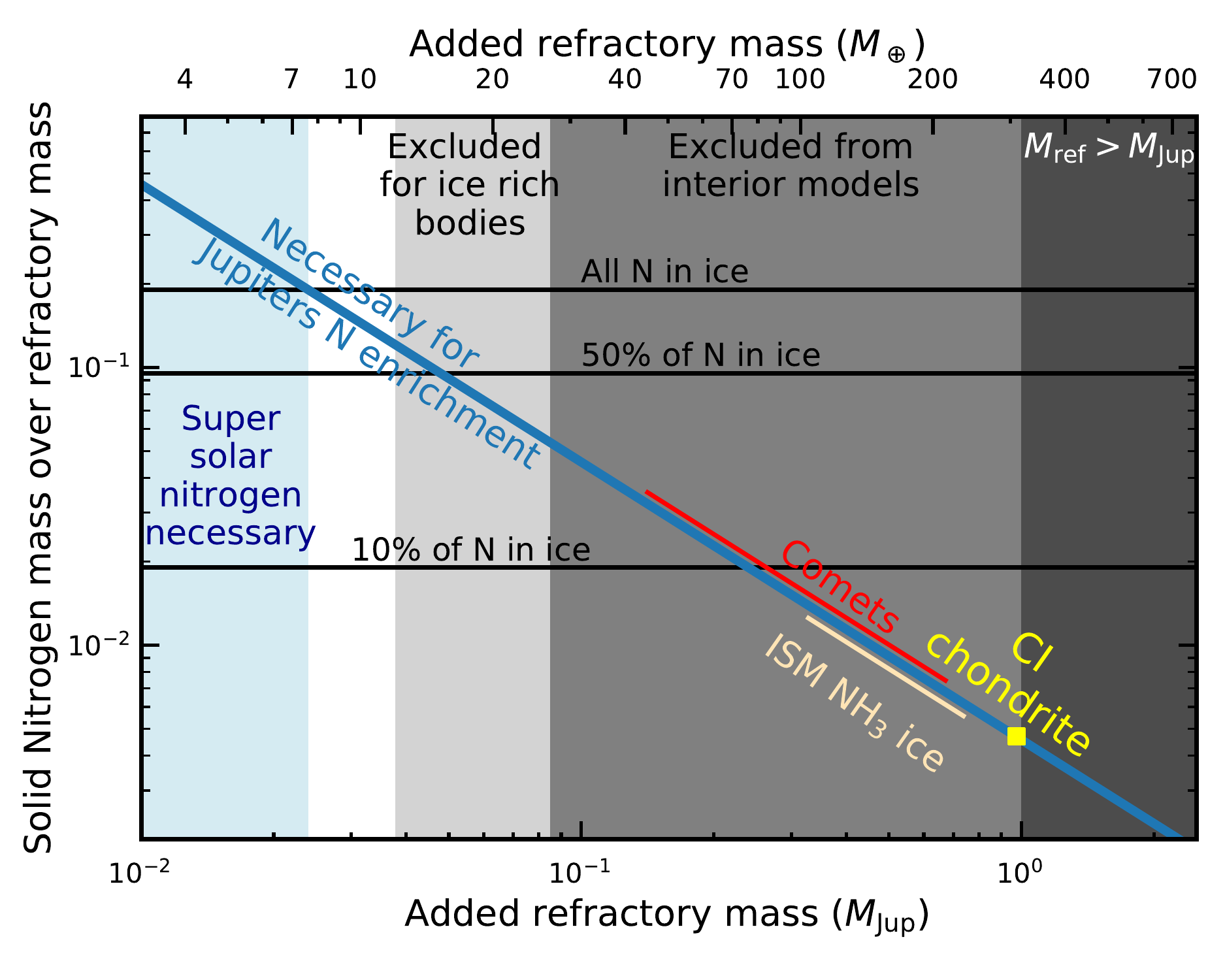}
\caption{\label{fig:nitrojup}Nitrogen to refractory mass ratio of enriching solids necessary to reach Jupiter's Nitrogen content as a function of total refractory mass added to Jupiter. It is assumed that Jupiter has accreted solar composition gas and that the excess nitrogen was brought in frozen on solids. The upper limit to the total heavy element mass is taken to be 27 $M_\oplus$ \citep{Wahl2017}. This can be in the form of ice poor or ice rich bodies, for ice poor bodies the refractory mass is assumed to be the total mass of heavy elements, for ice rich bodies, assuming \ce{H2O} and \ce{CO} are frozen out, as would be expected for a very \ce{N2} rich body, the heavy element mass is around two times the refractory mass, meaning that only $\sim$12 $M_oplus$ of refractories can be accreted. 
The nitrogen to refractory mass ratio of ISM grains without \ce{N2} ice \citep{Boogert2015}, comets \citep{Altwegg2019} and CI chondrites \citep{Lodders2009} have been added for comparison. None of these have a high enough nitrogen fraction to enhance Jupiter's atmosphere. }
\end{figure*}

\edt{
In Fig.~\ref{fig:nitrojup} we show this relation along with mass constraints based on the known properties of Jupiter and the solar system. The relative solid nitrogen mass and refractory mass is based on the nitrogen-to-refractory mass ratio, assuming that the volatile and refractory elemental abundances add up to the proto-solar abundance of \citet{Asplund2009}. We assume that the refractories contain all of the sulfur, iron, magnesium and silicon available of the ISM. Furthermore, we assume that the silicon is in equal parts enstatite and forsterite \citep[e.g.][]{Meeus2009}, resulting in an average of 3.5 oxygen atoms per silicon in the refractories. Finally we follow \citet{PontoppidanPPVI} and \citet{Bergin2015} and place 25\% of the total carbon in the refractory phase. This leads to a gas-to-refractory mass ratio of 186 and a nitrogen-to-refractory mass ratio of 0.2, the effective upper limit to the amount of nitrogen that can be added to the solid-phase. Above this mass ratio, the outer disk would have had to be more enhanced in nitrogen than seen in the solar photosphere, a scenario we find unlikely. In the outer parts in the disk, where all non-noble elements heavier than hydrogen are in the solids, the total gas-to-solid ratio drops to 77, so the total solid mass in these regions is about twice the available refractory mass. 

The gravitational moments measured by \textit{Juno} \citep{Bolton2017,Folkner2017,Iess2018} limit the total mass of heavy elements in the planet to be between 24 and 27 Earth masses \citep{Wahl2017}. Hence we can immediately neglect any refractory source that would need to accrete masses $\gtrsim 27$ M$_\oplus$. Furthermore, assuming that Jupiter accretes near the \ce{N2} snowline, the pebbles that are incorporated into the planet will also carry a significant portion of \ce{H2O} and CO ice.
}

\edt{The combination of the accreted heavy element mass and the available nitrogen} strongly limit the amount and composition of the solids that have enriched Jupiter's atmosphere. \edt{In the case of ice-free bodies} this requires bodies that $>25\%$ of the available nitrogen budget is incorporated in these bodies. \edt{However, for ice-rich bodies,} the total heavy element mass \edt{(which includes the ice)} is about twice the refractory mass, which further rules out the enrichment of Jupiter by nitrogen poor bodies and moves the minimum solid nitrogen fraction \edt{required} to more than $\sim 75$\% of the total available nitrogen. Comets and Meteorites are strongly ruled out as carriers of the nitrogen enhancement of Jupiter's atmosphere, as they contain far too little nitrogen \citep{Lodders2009, Altwegg2019}. Since we require at least 75\% of the proto-solar nitrogen budget to be in the solid phase, to explain the \ce{N}-enrichment of Jupiter's atmosphere, it must have accreted this mass in the form of \ce{N2} ice.

This low temperature origin of the building blocks of Jupiter was also proposed by \citet{Owen1999, Owen2006}. However, as new observations have constrained the total amount of enriching solids, we need the majority of the \ce{N2} to be in the ice, and thus temperatures below 20 K \citep{Bisschop2007}, instead of a smaller fraction of trapped \ce{N2} in a water-rich ice, in which case \ce{N2} can be trapped in the ice up to temperatures of 40 K \citep{Lunine1985,Collings2004}.

\section{Implications of Jupiter's Nitrogen enrichment}

\subsection{Enrichment during formation}
The high nitrogen abundance in Jupiter necessitates the accretion of \ce{N2} rich gas or solids. Enriching Jupiter during its formation means that the nitrogen was from a local source. The accretion of very nitrogen enriched gas can only happen just inside the \ce{N2} ice line, at 60 AU (Fig.~\ref{fig:form_scenario}, A). At the same time, enrichment of the atmosphere by accretion of small bodies necessitates \ce{N2} ice to be present, which similarly requires early atmosphere growth outside of the \ce{N2} ice line (Fig.~\ref{fig:form_scenario}, B). Finally nitrogen outgassing from the core would necessitate a \ce{N2} rich core and thus core formation outside of the \ce{N2} ice line. The exact location of the \ce{N2} ice line in the early solar system is hard to constrain and estimates of the \ce{N2} ice line in protoplanetary disks around solar mass stars vary greatly, ranging between 20 and 80 AU \citep{Huang2016,Terwisga2019,Qi2019}. 

Forming Jupiter's core of around 10 Earth masses \citep[e.g.][]{Lambrechts2017}, at these radii is very hard to do by planetesimal accretion \citep{Bitsch2015} and would point at a pebble accretion or gravitational instability origin for Jupiter. 

Pebble accretion seems especially promising as building a core from pebbles would leave the \ce{N2} on the pebbles until it is captured within the gravitational influence. Models by \citet{Bitsch2019} show that, as long as the pebble flux is high enough in the outer disk, it is possible to form a cold Jupiter starting core growth as far out as 50 AU. Taking an optimistic estimate for both the pebble accretion efficiency \citep[10\%][]{Ormel2018} and the total mass of pebble accreted onto the proto-Jupiter (7.5 $M_\oplus$), indicates a pebble reservoir of 75 $M_\oplus$ of refractories, or equivalently 0.05 $M_\odot$ of gas, outside the \ce{N2} iceline. This translates in to disks with radii between 30 and 100 AU, assuming a surface density power law slope between -1 and -0.5 \citep{Tazzari2016}. The total amount of dust necessary to form these pebble in the model is even larger \citep{Bitsch2019}, indicating an even larger and more massive disks would be necessary. This puts the proto-solar nebula among the largest and most massive disks currently observed \citep{Tazzari2016,Terwisga2019}.

In the case of formation by a gravitational instability of the gas disk, the energy released by the collapse of gas would locally heat the disk and evaporate the \ce{N2} ice off the grains, making accretion of N-enhanced material difficult \citep{Ilee2017}. The large transport rates in a gravitational unstable disk would quickly smooth any pre-exisiting overdensities in the gas or dust, making it difficult to build a nitrogen enriched object \citep{Kratter2016}. In all of the cases discussed above, \ce{N2} needs to be frozen out in the part of the disk where Jupiter is forming. This indicates that the disk needs to be large and cold, and thus likely to be in either the late class I or early class II stage, as the younger, still embedded disks are too warm to have CO, and thus \ce{N2} frozen out \citep{vantHoff2018}.

\subsection{Enrichment after formation}

It could also be possible to enrich Jupiter's atmosphere after it formed and has accreted the majority of its atmosphere. At this point we require the N-enriched bodies to be formed at large radii, while Jupiter is formed at smaller radii, for example the water ice line. The enriching bodies in this case need to be large, roughly kilometres in size, as they need to be able to hold-on to their \ce{N2} while traveling to Jupiter. If enrichment happens in the disk stage, these bodies need to be big enough not be be trapped in the pressure maximum caused by Jupiter, but not too big. As \ce{N2} is very volatile, and any internal heating by large impacts, or though radioactive decay \citep[e.g.][]{Prialnik1987} will lead to lower \ce{N2} abundances in the solids.

This scenario requires a very strict set of circumstances: assuming a core of around 10 $M_\oplus$ is needed to start gas accretion \citep{Lambrechts2017}, then this leaves at most 15 $M_\oplus$ of heavy elements, that is ice and refractories, that can be added. The minimal amount of refractories needed, assuming it manages to capture all the available \ce{N2} is 7.5 $M_\oplus$.  If Jupiter formed at the water ice line, there is a part of the disk ($\sim$15 AU wide) that does not contribute to the enrichment in Jupiter's atmosphere, while a significant amount of mass from the outer disk ($\gtrsim 20$ AU) makes it to Jupiter with its volatile component intact. This seems highly unlikely - which further argues for a young Jupiter forming very close, or even beyond the \ce{N2} ice line.

    
\section{Discussion}

\subsection{Nitrogen rich bodies in the solar system}
Up until now, there is little evidence of bodies incorporating the bulk of the proto-solar \ce{N2} as ice in the solar system \citep{Glein2018}. Pluto might have incorporated a significant amount of \ce{N2} ice at its formation, but without a measurement of the nitrogen isotopic ratio its origin is open to speculation \citep{Mandt2017}. Finding bodies that incorporated and still contain a significant fraction of the primordial nitrogen would point to the possible reservoir of Jupiter enriching bodies, and their current orbits could be indicative of the formation location of Jupiter. Both the very \ce{CH3OH} rich 2014 MU$_{69}$ \citep{Stern2019} and the comet C/2016 R2, which has a high \ce{N2}/CO ratio measured \citep{Opitom2019} could be one of these bodies. This indicates that bodies rich in \ce{N2} exist outside the orbit of Neptune.

The \ce{^{14}N/^{15}N} nitrogen isotopic ratio can be used to look at the origin of nitrogen in other bodies as well. There is a large discrepancy between the solar nitrogen isotopic ratio and the isotopic ratio found in many comet \citep[e.g.][]{Mumma2011}. This is most likely due to fractionation processes either in the ISM or in the proto-planetary disk enriching \ce{HCN} and \ce{NH3} and derivatives in \ce{^15N} \citep{Terzieva2000, Visser2018}. As these species are less volatile than \ce{N2}, ices above the sublimation temperature of \ce{N2} can easily be enriched in \ce{^15N}.

The nitrogen isotopic ratio in Jupiter is the same as the one measured in the solar wind, implying that Jupiter's nitrogen is indeed coming from the bulk nitrogen reservoir of the proto-solar nebula \citep{Fletcher2014}. Saturn has a similar nitrogen isotopic ratio to Jupiter as well as a similar overabundance of total nitrogen over the sun, indicating that Saturn likely inherited its nitrogen from the same source as Jupiter \citep{Fletcher2014, Atreya2016}. Other bodies which have measured isotopic ratios are up to a factor three lower than the solar value \citep{Niemann2010, Mandt2014, BockeleeMorvan2015, Mandt2017}, which includes, meteorites, comets and Titan. Indicating that these bodies did not accrete their nitrogen from the bulk \ce{N2} reservoir. 

\subsection{Carbon and oxygen in Jupiter}

\begin{table*}[]
    \caption{\label{tab:O/H+C/H} Carbon and oxygen abundances relative to solar predicted for Jupiter's atmosphere assuming different contributing sources.}
    \centering
    \begin{tabular}{l c c c}
    \hline
    \hline
    Incorporated species  & [O/H]/[O/H]$_\star$ & [C/H]/[C/H]$_\star$ & C/O \\
    \hline
    \multicolumn{4}{c}{Accretion of \ce{N2} enriched gas}\\
    1: Gaseous \ce{N2} and \ce{CO}  & 1.8 & 3.6 & 1.1\\
    2: 1 + refr. C and \ce{H2O} ice & 2.7 & 4.0$^{*}$ & 0.8 \\
    3: 2 + silicates & 3.3 & 4.0$^{*}$& 0.66\\
    \hline 
    \multicolumn{4}{c}{Accretion of \ce{N2} rich pebbles}\\
    1: \ce{N2}, \ce{CO} and \ce{H2O} ice & 3.6 & 3.6 & 0.55 \\
    2: 1 + refractory carbon & 3.6 & 4.5 & 0.7\\
    3: 2 + silicates & 4.5 & 4.5 & 0.55\\
    \hline
    \end{tabular}
    \tablefoot{$^*$Set to match Jupiter's carbon abundance. }   
\end{table*}

Working with the assumption that Jupiter did not accrete a significant amount of solids after accreting most of its gas, it is possible to use the different formation scenarios and measured carbon content in the planet, to predict the oxygen content of Jupiter. These prediction depend critically on what is assumed to happen with the refractory carbon \citep[here 25\% of total carbon][]{PontoppidanPPVI,Bergin2015} and oxygen contained in refractories. Assuming silicates are in a 50--50 mix of \ce{SiO3} and \ce{SiO4} ions and iron not in the form of iron oxides, about 23\% of the oxygen is refractory \citep{Costantini2005,Meeus2009}. For simplicity we assume that all available volatile carbon is in CO, which then contains 40\% of the total oxygen and the remaining 37\% of the oxygen in \ce{H2O}, representative of gas in the inner regions of disks \citep{PontoppidanPPVI}. As such we are ignoring the few to tens of percent of carbon that can be contained within \ce{CO2} in the ice \citep{PontoppidanPPVI, Boogert2015, LeRoy2015}. 

Table.~\ref{tab:O/H+C/H} shows the oxygen and carbon abundances in Jupiter as predicted from different enrichment scenarios. In the case that \ce{N2} is accreted from \ce{N2} enriched gas, we assume that that gas is also enriched in \ce{CO} by the same factor, which leads to an enrichment in both oxygen and carbon. The extra carbon cannot explain the full carbon enrichment observed in Jupiter's atmosphere and thus additional carbon from the refractory reservoir is necessary. Here one can assume that only the water ice on these grains enriches the atmosphere in oxygen, or that both the water ice and the silicates deposit oxygen in the atmosphere. In all cases a super solar C/O ratio is found. 

In the case that excess \ce{N2} was accreted as solid \ce{N2} on top of a 1 $M_\mathrm{Jup}$ solar composition atmosphere, it is safe to assume that all of the ices on the grains would also enrich the atmosphere. Again this does not match the carbon enrichment in the atmosphere and an additional carbon source is necessary. However, including all of the refractory carbon that is brought in by these grains would raise the carbon abundance in Jupiter to a value higher than the nominal measured value, but still within the range of observations. Finally adding enrichment by the full release of oxygen from the silicates brings the oxygen enhancement up to 4.5 times solar and the C/O ratio back to solar. Hence a solar or slightly super solar C/O ratio is predicted for Jupiter. 

\section{Conclusions}
The nitrogen enrichment in Jupiter's atmosphere makes it likely that Jupiter formed at much larger radii than it is observed now. At these radii, core formation due to pebble accretion onto a planetesimal seems to be the most likely scenario as it would naturally bring in a lot of nitrogen rich ice. This would however, necessitate a cold, massive and large disk to have a massive enough \ce{N2} ice reservoir to enrich Jupiter and enough pebbles flowing in these cold regions to be able to form Jupiter. The proto-solar nebula should thus have looked like the most massive and largest proto-planetary disks that are currently observed. 

This formation scenario necessitates the formation of Jupiter's core at a time that the disk was cool enough to have \ce{N2} as an ice. Furthermore, the mass of pebbles necessary to enrich Jupiter's atmosphere imply that formation of Jupiter in a large, massive disk. Implying that the proto-solar disk was analogous to the largest proto-planetary disks currently observed. Jupiter's atmosphere should be enriched in oxygen, in this case, with a O/H below 4.5 times solar, with the preferred models predicting O/H between 3.6 and 4.5 times solar.

\begin{acknowledgements}
The authors thank the referee for a constructive report that improved the quality of the paper. Astrochemistry in Leiden is supported by the Netherlands Research School for Astronomy (NOVA). This project has made use Matplotlib \citep{Hunter2007}.
\end{acknowledgements}
\bibliographystyle{aa}
\bibliography{Lit_list}

\end{document}